\documentclass[12pt]{article}
\newcommand{\p}[1]{(\ref{#1})}
\newcommand{\be}{\begin{equation}}
\newcommand{\ee}{\end{equation}}
\newcommand{\bea}{\begin{eqnarray}}
\newcommand{\eea}{\end{eqnarray}}

\begin{document}
\title{$N=4$ supersymmetric Eguchi--Hanson sigma model in $d=1$}

\setcounter{page}{1}
\maketitle
\begin{center}
\v{C}. Burd\'{\i}k${}^1$, S. Krivonos${}^2$, A. Shcherbakov${}^2$ \vskip 0.5 cm
\small ${}^1$Department of Mathematics, Czech Technical University, Trojanova 13,  \\ 120 00 Prague 2 \vskip 0.3 cm
${}^2$Bogoliubov Laboratory of Theoretical Physics, JINR, 141980 Dubna, Russia
\end{center}

\begin{abstract}
We show that it is possible to construct a supersymmetric mechanics with four supercharges possessing not
conformally flat target space. A general idea of constructing such models is presented.
A particular case with Eguchi--Hanson target space is investigated in details: we present the standard and
quotient approaches to get the Eguchi--Hanson model, demonstrate their equivalence, give a full set of
nonlinear constraints, study their properties and give an explicit expression for the target space metric.
\end{abstract}

\section{Introduction}

In many studies of $N=4$ supersymmetric mechanics (SM) (see, e.g. \cite{{sm1},{sm2},{sm3},{sm4}}) actions invariant under
extended supersymmetries have been constructed in terms of components or $N=1$ superfields. The requirement of invariance
with respect to additional supersymmetries (non-manifest ones) puts some restrictions on the relevant target geometries.
For a long time it has been almost evident that such an approach gives the most general types of $N=4$ $d=1$ supersymmetric
sigma models, at least for the $N=4$ hypermultiplet. The main argument for such a statement is the property of $N=4$ $d=1$
hypermultiplet which contains no
auxiliary fields. Therefore, the formulation in terms of unconstrained superfields, being constructed, should reproduce the same
component actions and it seems that there is no place for novel features. Of course, it is quite desirable to have a formulation
of a SM in an appropriate superspace where all its supersymmetries are manifest and off-shell. Such formulations have been
pioneered in \cite{{leva},{is},{ikp},{bp}} and further elaborated in, e.g. [10-19]. The actions have been constructed
in the standard type of $N=4$ $d=1$ superspaces as well as in $N=4$ $d=1$ harmonic superspaces. But in all cases the models reveal the
same structure of their bosonic target spaces -- conformal flatness (with some additional restrictions).

Being quite general, these results keep opened only one way out of the conformally flat type metrics of the target spaces of $N=4$ SM --
to use {\it nonlinear} supermultiplets.
The first step in this direction has been done in \cite{bbkno}, where a new variant of $N=4$ SM with {\it nonlinear} supermultiplet proposed in
\cite{15} was constructed. The target space metric in this model is still conformally flat and K\"{a}hler one, but
it is not a special K\"{a}hler one as one may expect from standard consideration based on the {\it linear} supermultiplets.
Of course, the main question is how to find the proper constraints describing the {\it off-shell nonlinear} $N=4$ supermultiplets.
A possible answer is proposed in \cite{DK}. The idea is to use the reduced version of the equations of motion
for the $N=2$, $d=4$ hypermultiplets as the constraints defining a {\it nonlinear} $N=4$ supermultiplet. The reduction applied in
\cite{DK} consists in discarding space-time indices of $N=2$, $d=4$ spinor covariant derivatives and getting
$N=4$ supermultiplets as a result. Moreover, these supermultiplets are off-shell in $d=1$. In \cite{DK} only the case of
Taub-NUT sigma model is considered. In the present letter we will go further and will demonstrate that another
interesting example of $N=4$ $d=1$ sigma model with Eguchi-Hanson (EH) target space metric can be constructed in a
similar way. Moreover, we will demonstrate that the powerful quotient construction \cite{qu} which is widely used in $N=2$, $d=4$
harmonic superspaces (see e.g.\cite{iv}) works quite well in $N=4$ $d=1$ case.

\section{N=4 Eguchi-Hanson sigma model: the standard formulation}
The main purpose of this letter is to give an action for an $N=4$ SM with the Eguchi-Hanson manifold as its
target space. The first step is to define the corresponding $N=4$ $d=1$ {\it nonlinear} supermultiplet.
The point of departure is the $N=2$, $d=4$ sigma model with EH metric \cite{qu}. It may be described by the following
action in $N=2$, $d=4$ harmonic superspace \cite{bible}:
\be\label{n2d4EH}
S_{EH} \sim \int {\,}\mbox{d}\zeta^{(-4)}{\,}\mbox{d}u \left[ \left( {\cal D}{}^{++}\omega\right)^2 - \frac{ \left( \lambda^{++}\right)^2}{\omega^2}
\right],
\ee
where dimensionless quantity $\lambda^{++}$ is given by
\be\label{lambda}
\lambda^{++}=\lambda^{ij}u^+_i u^+_j
\ee
in terms of the real isovector coupling constant $\lambda^{ij}$.
The equation of motion that follows from (\ref{n2d4EH}) is
\be\label{EH1}
\left( {\cal D}{}^{++} \right)^2 \omega = \frac{\left( \lambda^{++}\right)^2}{ \omega^{3}}.
\ee
Another important condition is an analyticity of the superfield $\omega$:
\be\label{an}
{\cal D}^+_\alpha \omega = {\overline{\cal D}}{}^+_{\dot\alpha} \omega=0 \;.
\ee
Now we transfer the constraints (\ref{EH1}) and (\ref{an}) in $N=4$ $d=1$ harmonic superspace:
\bea
\left( D^{++} \right)^2 \omega = \frac{\left( \lambda^{++}\right)^2}{ \omega^{3}}, \label{eh1} \\
 D^{+a} \omega = 0 \;. \label{eh2}
\eea
Here, harmonic derivative $D^{++}$ is defined as (in the central basis)
\be\label{hD}
D^{++} = u^{+i}\frac{\partial}{\partial u^{-i}},
\ee
while the spinor derivatives $D^{\pm a}$ are obtained from the standard derivatives in $N=4$ $d=1$ superspace
${\bf R}^{1|4}$
\be
{\bf R}^{1|4}=\{t,\theta_{ia}\; ; a,i=1,2 \}, \qquad \bar t = t, \quad
  \overline{\rule{0pt}{0.9em}\theta_{ia}}=\theta^{ia}=\varepsilon^{ij}\varepsilon^{ab}\theta_{jb}\nonumber
\ee
by taking their projections onto the harmonics
\be
  D^{\pm\,a}=u^\pm_i D^{ia}=\mp\frac{\partial}{\partial\theta^\mp_a}+i\theta^{\pm\,a}\partial_t, \qquad
  \theta^{\pm\,a}\equiv\theta^{ia}u^{\pm}_i.
\ee

The key difference of the constraints (\ref{eh1}), (\ref{eh2}) from their four dimensional counterpart (\ref{EH1}), (\ref{an})
is that they describe {\it off-shell nonlinear} $N=4$ $d=1$ supermultiplet with four bosonic and four fermionic
components. The simplest way to see this is to consider a limit case $\lambda^{++}=0$ when the constraints
(\ref{eh1}), (\ref{eh2}) can be rewritten as follows \cite{bible}
\be\label{q0}
\omega= u^{-}_i q^{+i}=u^{-}_1 {\bar q}{}^+ - u_2^- q^+\, \Rightarrow \, D^{++}q^+=0, \quad D^{+a}q^+ =0.
\ee
The constraints (\ref{q0}) on the superfields $q^+$ are just standard ones defining $N=4$ $d=1$ hypermultiplet in harmonic superspace \cite{{16},{15}}.
The nonlinearity does not change the structure of the supermultiplet. It simply makes the transformation properties
of the components highly nonlinear.

Finally, we are ready to present the action for $N=4$ $d=1$ SM with EH metric in its target space. The action is
\be\label{action1}
S_{EH} \sim \int {\,}\mbox{d}\zeta^{(-2)}{\,}\mbox{d}u \; \omega D^{++} \dot\omega \;,
\ee
where the $N=4$ superfield $\omega$ is subjected to constraints (\ref{eh1}), (\ref{eh2}). Of course, the integration in (\ref{action1})
goes over $N=4$ $d=1$ analytic superspace ${\bf AR}^{1+2|2}$ \cite{16}
\be\label{ah}
{\bf AR}^{1+2|2}=\{ \zeta, u\}=\{t_A=t+i\theta^{+a}\theta^-_a,\theta^{+a},u^+_i, u^-_i\}.
\ee

In the next section we will pass to the components to demonstrate explicitly that the action (\ref{action1})
describes the desired $N=4$ SM.

In what follows we will use a new complex spinor coordinate $\theta^{+}$ and its conjugated ${\bar\theta}^{+}$
(in the sense of tilda-conjugation \cite{bible}, though it is denoted through the bar sign)
$$\theta^{+}=\frac 1{\sqrt{2}}\left(\theta^{+1}+i\theta^{+2}\right), \quad
  {\bar\theta}^{+}=-\frac i{\sqrt{2}}\left(\theta^{+1}-i\theta^{+2}\right)$$
with the following conjugation properties
$$\widetilde{\theta^{+}}={\bar\theta}^{+},\quad \widetilde{{\bar\theta}^{+}}=-\theta^{+}.$$
In terms of these coordinates the harmonic derivative (when acts on an analytic superfield) gets the form
$$D^{++}= u^{+\,i}\frac{\partial\phantom{u}}{\partial u^{-\,i}}-2i\theta^{+}{\bar\theta}^{+}\frac{\partial}{\partial t}.$$

\section{N=4 Eguchi-Hanson sigma model: an alternative formulation}
Similarly to the four dimensional case \cite{qu} one may directly solve the equation \p{eh1} and substitute its solution
into the action \p{action1}. But it is preferable to use an equivalent form of the action \p{action1} written in terms
of two hypermultiplets. The action we are going to consider resembles the quotient method action \cite{qu} and reads
\be\label{act}
S = - i\int {\,}\mbox{d}t {\,}\mbox{d}\zeta^{(-2)} \Biggl[ Q^{+a}\dot{\bar Q}^+_a + A\left( \frac i2 Q^{+a}\bar Q^+_a - \lambda^{++}\right)\Biggr],
\qquad \bar Q^+_a\equiv \widetilde{Q^{+a}}
\ee
along with nonlinear constraints
\be\label{constr}
D^{++} Q^{+a} = i \xi^{++} Q^{+a}, \qquad D^{++} \bar Q^{+a} = - i \xi^{++} \bar Q^{+a}, \qquad \widetilde{\xi^{++}}=\xi^{++}.
\ee
Here $Q^{+a}(t,\theta^{+},{\bar\theta}^{+},u^{\pm})$ is a complex analytic superfield
(analog of the Fayet--Sohnius hypermultiplet in four dimensions),
$\xi^{++}(t,\theta^{+},{\bar\theta}^{+},u^{\pm})$ is a real analytic superfield and $A(t,\theta^{+},{\bar\theta}^{+},u^{\pm})$ is a Lagrange
multiplier maintaining the quadratic constraint.

Let us stress that the nonlinear constraints \p{constr} are off-shell ones. They describe the supermultiplet with
eight real bosonic and eight real fermionic degrees of freedom. Therefore, one should somehow reduce the number of the components
in \p{act} to $4^b + 4^f$ which are present in \p{action1}. This may be done due to $U(1)$ gauge invariance of the action \p{act}
and constraints \p{constr}. Indeed, one may check that the action \p{act} is invariant under
\be\label{sym}
Q^{+a}{}'=e^{i\alpha}Q^{+a}, \qquad {\bar Q}^{+a}{}'=e^{-i\alpha}{\bar Q}^{+a}, \qquad A' = A + 2\dot\alpha
\ee
where gauge parameter $\alpha$ is a real analytic function. To have constraints (\ref{constr}) invariant
the superfield $\xi^{++}$ should transform as
\be\label{xi}
\xi^{++}{}'=\xi^{++}+D^{++}\alpha.
\ee

Now we can demonstrate that the action \p{act} supplemented by the constraints \p{constr} is equivalent
to the action \p{action1} with constraints \p{eh1}. First of all we represent the superfields $Q^{+a}, {\bar Q}{}^{+a}$
in the form
\be\label{rel}
Q^{+a}=u^{+a} \omega + u^{-a} f^{++}, \qquad \bar Q^{+a}=u^{+a} \bar \omega + u^{-a} \bar f^{++}.
\ee
Due to constraints (\ref{constr}) we have
\be\label{const}
D^{++}\omega + f^{++} = i \xi^{++} \omega, \qquad
D^{++}f^{++} = i\xi^{++}f^{++}.
\ee
Therefore, using the first equation in \p{const}, we may express $Q^{+a}$, $\bar Q^{+a}$ in terms of $\omega$ and $\bar\omega$ respectively
\be
Q^{+a}=u^{+a} \omega - u^{-a} (D^{++}\omega - i \xi^{++} \omega), \qquad
\bar Q^{+a}=u^{+a} \bar \omega - u^{-a} (D^{++}\omega + i \xi^{++} \omega).
\ee
The second equation gives rise to a constraint
\be\label{con}
(D^{++})^2 \omega - i \omega D^{++}\xi^{++} - 2i \xi^{++}D^{++}\omega - (\xi^{++})^2 \omega = 0.
\ee
Rewriting the action (\ref{act}) in terms of $\omega$ hypermultiplet we will get
\bea\label{omega_act}
S&=&\int {\,}\mbox{d} u {\,}\mbox{d}\zeta^{(-2)} \left[ -\omega D^{++}\dot{\bar\omega} + \dot{\bar\omega}D^{++}\omega
  - i\xi^{++} \partial_t (\omega\bar\omega) \right. \nonumber\\
&&\left. + A \left(\frac i2\left(\bar\omega D^{++}\omega - \omega D^{++}\bar\omega \right)
  +\xi^{++}\omega\bar\omega - \lambda^{++}  \right) \right].
\eea
The gauge symmetry \p{sym} realized on $\omega$ and $\bar\omega$ as
\be
\delta\omega = i \alpha \omega, \qquad \delta\bar \omega = - i \alpha \bar\omega
\ee
gives the possibility to require that $\omega$ be real
\be\label{gauge1}
\omega=\bar\omega.
\ee
In this gauge
variation with respect to the Lagrange multiplier $A$ leads to the following expression for $\xi^{++}$
\be\label{xi_sol}
\xi^{++}=\frac{\lambda^{++}}{\omega^2}
\ee
and the action (\ref{omega_act}) acquires the form (\ref{action1}) with $\omega$ being restricted according to equation (\ref{eh1}).
Thus, the action \p{act} along with the constraints \p{constr} is equivalent to the action \p{action1} with additional constrains \p{eh1}.

\section{N=4 Eguchi-Hanson sigma model: Wess-Zumino gauge}
In close analogy with four dimensional case \cite{qu} one may choose another gauge instead of \p{gauge1}.
Indeed, the transformation properties \p{xi} of the $\xi^{++}$
allows us to eliminate all but one real degree of freedom in $\xi^{++}$
\be\label{WZ}
\xi^{++}=i\theta^{+}{\bar\theta}^{+} V(t).
\ee
This remaining component $V(t)$ still transforms under a residual $U(1)$ gauge transformation \p{sym} as
\be
\delta V=-2\dot\alpha(t)
\ee
with $\alpha(t)$ being now a function of $t$ only.

In the gauge \p{WZ} the constraints (\ref{constr}) can be easily solved. The solution
restricts the hypermultiplet components
\be\label{Q}
Q^{+a}=q^{+a}+\theta^{+} \psi^a + {\bar\theta}^{+} \chi^a + i\theta^{+}{\bar\theta}^{+} P^{-a}, \qquad
\bar Q^+_a=\bar q^+_a+\theta^{+} \bar \chi_a - {\bar\theta}^{+} \bar \psi_a + i\theta^{+}{\bar\theta}^{+} \bar P^-_a
\ee
to have the following form
\bea
&&q^{+a}(t,u^\pm)=q^{ia}(t)u^+_i, \qquad \bar q^+_a(t,u^\pm)=\bar q_{ia}(t)u^{+i},\nonumber\\
&&\psi^a(t,u^\pm)=\psi^a(t), \phantom{{}^i u^+}\qquad \chi^a(t,u^\pm)=\chi^a(t),\\
&&P^{-a}(t,u^\pm)=(2\dot q^{ia}+i V q^{ia} )u^-_i, \qquad
  \bar P^-_a(t,u^\pm)=(2\dot{\bar{q}}_{ia}-iV\bar q_{ia} )u^{-i}.\nonumber
\eea
To get the components form of the action \p{act} one should also solve the
quadratic superfield constraint which is encoded there.
These constraints restrict the bosonic components $q^{ia}$ and $\bar q_{ia}$ of the hypermultiplet by three conditions
\be\label{q}
\frac i2 q^{(ia} \bar q^{j)}_a =\lambda^{(ij)},
\ee
and express a half of the spinor degrees of freedom through the other
\be\label{chi}
\chi^a=2\;\frac{\bar q^a_i q^{ib}\bar\psi_b}{\bar q^{jc}\bar q_{jc}}, \qquad
\bar\chi_a=-2\;\frac{q^i_a \bar q_{ib}\psi^b}{q^{jc}q_{jc}}.
\ee
In addition they allow to
get an explicit expression for the only component of $\xi^{++}$ in the Wess--Zumino gauge \p{WZ}
\be\label{V}
V=\frac{i(\dot q^{ia}\bar q_{ia}-q^{ia}\dot{\bar q}_{ia})+\psi^a\bar\psi_a+\chi^a\bar\chi_a}{q^{jb}\bar q_{jb}}.
\ee
Let us note, that the above expressions possess the proper transformation properties with respect to the residual $U(1)$ gauge symmetry.

Finally, performing Grassmann integration and integration over harmonics $u^\pm_i$, action (\ref{act}) acquires the following form
\bea\label{act2}
S=\int {\,}\mbox{d} t \Biggl[ -2\dot q^{ia}\dot{\bar q}_{ia}-\frac12\;\frac{(\dot q^{ia}\bar q_{ia}-q^{ia}\dot{\bar q}_{ia})^2}{q^{jb}\bar q_{jb}}
  +i\psi^a\dot{\bar\psi}_a+i\chi^a\dot{\bar\chi}_a\Biggr.&&\nonumber\\
\qquad\qquad\Biggl. +\frac i2 (\psi^a\bar\psi_a+\chi^a\bar\chi_a)\frac{\dot q^{ia}\bar q_{ia}-q^{ia}\dot{\bar q}_{ia}}
  {q^{jb}\bar q_{jb}}\Biggr]&&
\eea
with $q^{ia}$ and $\chi^a$ satisfying (\ref{q}) and (\ref{chi}). The bosonic part of the action \p{act2} together with the
constraints \p{q} are just the one dimensional
version of the Lagrangian of ref. \cite{CF}. Thus we conclude that the action \p{action1} with the constraints \p{eh1} as
well as the action \p{action1} with the constraints \p{constr} describe $N=4$ $d=1$ supersymmetric SM with Eguchi-Hanson
metric of its target space.

It is tempting to have the explicit form of the action \p{act2}, at least for its bosonic part. The only constraint we have yet to solve is
quadratic one (\ref{q}). To proceed we, first,
choose a specific parametrization of the $\lambda^{(ij)}$
\be
\lambda^{(ij)}=\left( \begin{array}{cc} 0 & i \lambda \\ i\lambda & 0 \end{array} \right), \qquad
  \overline{\rule{0pt}{1em}\lambda^{(ij)}} = \lambda_{(ij)}=\varepsilon_{ik}\varepsilon_{jl}\lambda^{(kl)}, \quad \bar\lambda=\lambda.
\ee
Then we may solve the constraints \p{q} as follow
\bea\label{Qpar}
q^{11} &=& \frac{\Lambda f(u)}{\sqrt{1+\Lambda{\rule{0pt}{1em}\bar\Lambda}}}e^{-\frac i2 \phi}, \qquad
  q^{12} = \frac{f(u)}{\sqrt{1+\Lambda{\rule{0pt}{1em}\bar\Lambda}}}e^{\frac i2 \phi},\\
q^{21} &=& \frac{h(u)}{\sqrt{1+\Lambda{\rule{0pt}{1em}\bar\Lambda}}}e^{-\frac i2 \phi}, \qquad
  q^{22} = - \frac{{\rule{0pt}{1em}\bar\Lambda} h(u)}{\sqrt{1+\Lambda{\rule{0pt}{1em}\bar\Lambda}}}e^{\frac i2 \phi},\nonumber
\eea
where
$$f(u)=e^{u/2}+\lambda e^{-u/2}, \qquad h(u)=e^{u/2}-\lambda e^{-u/2}.$$
Substituting all these into action \p{act2} and omitting all fermionic terms we will get
\bea\label{act_bos}
S_{bos}&=&\int {\,}\mbox{d} t \Biggl[ \left(e^u + \lambda^2 e^{-u} \right) \left( {\dot u}{}^2
  +\left(\dot\phi - i\frac{\Lambda{\dot{\bar\Lambda}}-\dot{\Lambda}{\rule{0pt}{1em}\bar\Lambda}}{1+\Lambda{\rule{0pt}{1em}\bar\Lambda}} \right)^2
  +\frac{4\dot{\Lambda}{\dot{\bar\Lambda}}}{(1+\Lambda{\rule{0pt}{1em}\bar\Lambda})^2}\right) \Biggr.\nonumber\\
&&\Biggl.-\frac{4\lambda^2}{e^u + \lambda^2 e^{-u}}\left( \frac{1-\Lambda{\rule{0pt}{1em}\bar\Lambda}}{1+\Lambda{\rule{0pt}{1em}\bar\Lambda}}\right)^2
  \left( \dot\phi+i\frac{\Lambda{\dot{\bar\Lambda}}-\dot{\Lambda}{\rule{0pt}{1em}\bar\Lambda}}{1-\Lambda{\rule{0pt}{1em}\bar\Lambda}}\right)^2 \Biggr].
\eea
One may explicitly check that the target space metric $G_{AB}$ corresponding to the action (\ref{act_bos})
$$
S_{bos}=\int {\,}\mbox{d} t\; G_{AB}(\Phi)\; \dot\Phi^A \dot\Phi^B, \qquad A,B=1,\ldots,4, \quad \Phi^A=(\Lambda,{\rule{0pt}{1em}\bar\Lambda},u,\phi)
$$
has vanishing Ricci tensor as it should be for a hyper K\"ahler metric.

\section*{Acknowledgments}
The authors thank Evgeny Ivanov for valuable discussions and remarks.
This work was partially supported by grant GACR 201/05/0857,
RFBR-DFG grant No 02-02-04002, grant DFG No 436 RUS 113/669, RFBR grant
No 03-02-17440 and a grant of the Votruba-Blokhintsev programme.

\end{document}